\documentclass[12pt]{article}

\usepackage{scicite}
\usepackage{graphics}

\usepackage{times}

\newenvironment{sciabstract}{%
\begin{quote} \bf}
{\end{quote}}

\newcounter{lastnote}
\newenvironment{scilastnote}{%
\setcounter{lastnote}{\value{enumiv}}%
\addtocounter{lastnote}{+1}%
\begin{list}%
{\arabic{lastnote}.}
{\setlength{\leftmargin}{.22in}}
{\setlength{\labelsep}{.5em}}}
{\end{list}}

\title{Observations of Plasmarons in a System of Massive Electrons}

\author
{OE Dial,$^{1\ast}$ RC Ashoori $^{1}$,LN Pfeiffer $^{2}$ KW West$^{2}$\\
\\
\normalsize{$^{1}$Massachusetts Institute of Technology, Cambridge, MA 02139}\\
\normalsize{$^{2}$Princeton, Princeton, NJ 08544}\\
\\
\normalsize{$^\ast$To whom correspondence should be addressed; E-mail:  dial@alum.mit.edu.}
}

\date{}

\begin{document} 


\baselineskip24pt


\maketitle


\begin{sciabstract}
Calculations of the single particle density of states (SPDOS) of
electron liquids have long predicted that there exist two distinct
charged excitations: the usual quasiparticle consisting of an electron
or hole screened by a correlation hole, and a ``plasmaron'' consisting
of a hole resonantly bound to real plasmons in the Fermi
sea\cite{Hedin67,Lundqvist67}. Using tunneling spectroscopy to
measure the SPDOS of a two-dimensional electronic system, we
demonstrate the first detection of the plasmaron in a system in which
electrons have mass.  We monitor the evolution of the plasmaron with
applied magnetic field and discover unpredicted ``magnetoplasmarons''
which appear in spectra as negative index Landau levels. These
sharp features corresponding to long-lived quasiparticles
appear at high energies where SPDOS structure is ordinarily 
broadened by electron-electron interactions.
\end{sciabstract}

The ``plasmaron'' was originally proposed by Hedin and Lundquist\cite{Hedin67,Lundqvist67} to exist in three-dimensional (3D) metals. 
However, aside from the unusual semi-metal Bismuth\cite{Tediosi07}, it has not been unambiguously observed in 3D, and some theories 
suggest that the 3D plasmaron should not exist\cite{Bergersen73}.  
In contrast, more recent theories predict a robust plasmaron in two-dimensional
electronic systems
(2DESs)\cite{vonAllmen92,Jalabert89a,Jalabert89b,Hwang08,Polini08}. Indeed,
the plasmaron has recently been observed in graphene, which contains a
2DES with massless electrons\cite{Bostwick10}.  We observe several key
differences between our observation of the plasmaron in system with
massive electrons and the experimental and theoretical results for
graphene.  Whereas in graphene the plasmaron is predicted to exist both above and
below the Fermi energy\cite{Hwang08} and exhibits a simple scaling law with electron
density, we find that in a system of massive electrons the plasmaron
exhibits a strong asymmetry about the Fermi energy and exhibits a
non-linear and unpredicted dependence on electron density.

We perform our tunneling measurements using time domain capacitance
spectroscopy (TDCS)\cite{Chan97,Dial07,Dial10}.  This technique allows
the measurement of the SPDOS with precisely calibrated energy and
density axes and minimal effects of heating.  In TDCS, the 2DES to be
studied is grown epitaxially inside of a capacitor.  One plate of the
capacitor, the tunnel electrode, is close enough to allow charge to
tunnel to and from the 2D system.  The other plate is distant enough
to be electrically isolated and is used to detect the tunnel current by
means of its image charge.  DC voltages can be used to tune the
density of the 2D system continuously, while tunnel voltages and
currents can be induced by applying voltage pulses across the
capacitor.  In the present measurements the 2DES is a 230\AA{} GaAs
quantum well.  The plates of the capacitor are 3D doped regions of
GaAs with an electron density of $1\times 10^{18} \mathrm{cm^{-2}}$.
The tunnel electrode is separated from the 2DES by a 130\AA{} tunnel
barrier of $x=.324$ $\mathrm{Al_xGa_{1-x}As}$ followed by 300\AA{} of
undoped GaAs which acts as a spacer layer to reduce
scattering\cite{Lebens87}.  This spacer is thin enough it is populated
by electrons due to the finite Thomas-Fermi screening length.  The
electrically isolated electrode is separated from the 2DES by 600\AA{}
of $x=.324$ $\mathrm{Al_xGa_{1-x}As}$.

\begin{figure*}
\resizebox{6in}{!}{\includegraphics{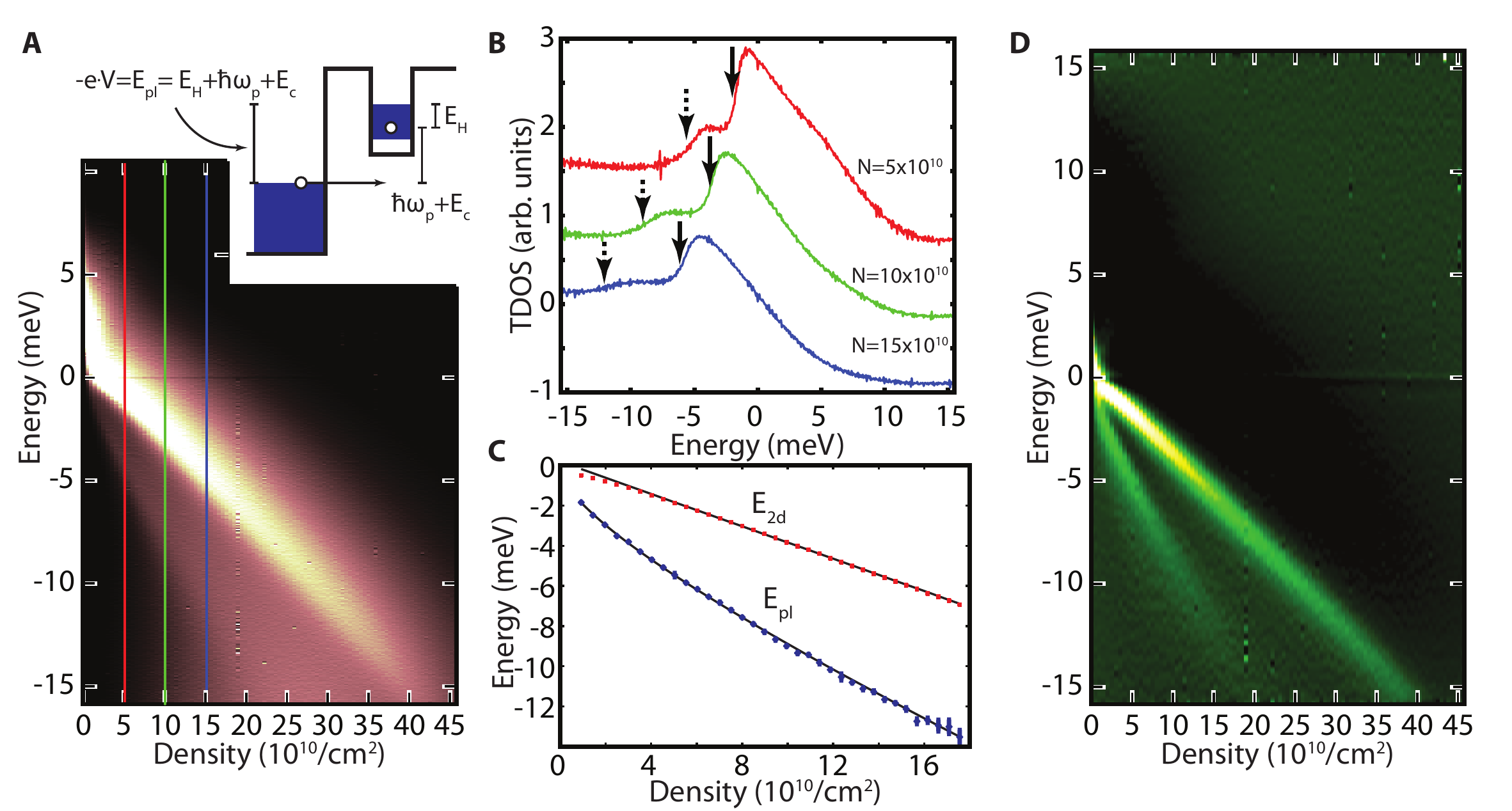}}
\caption{\label{zerofield}
TDCS spectra acquired at zero magnetic field display both a 2D band edge, and a second ``plasmaron'' edge at lower energies.  In \textbf{A}, the 2D band edge appears as a sharp peak due to tunneling matrix element effects.  The plasmaron is visible as a faint second step at more negative energies.  A small cartoon shows the measured energy of the plasmaron is negative, and roughly equal to the sum of the energy of the required hole, the required plasmon, and a coupling energy.  The two edges are indicated in the linecuts in \textbf{B} (offset vertically for clarity), with the band edge indicated with a
solid arrow, and the plasmaron edge indicated with a dotted arrow.  In \textbf{D}, the band edges can be emphasized by taking an extra derivative of the data along the energy axis ($d^2I/dV^2$).  Defining the ``edge'' as the peak in $d^2I/dV^2$, this allows the energies of the 2D band edge (red) and plasmaron band edge (blue) to be extracted, as shown in \textbf{C}.  Note the extreme asymmetry of the plasmaron band about the Fermi energy, ruling out inelastic scattering as a possible origin.}
\end{figure*}

Figure \ref{zerofield}a shows a typical TDCS spectrum acquired at 100
mK with zero magnetic field.  The horizontal axis is the density in
the 2DES, which is controlled by the DC bias across the device.  The
vertical axis is energy, with E=0 corresponding to the Fermi energy.
The band edge of the 2DES, corresponding to injecting a wavevector
$k=0$ hole, is visible as a abrupt edge that begins near $E=0$ at zero
density and moves downward in energy as the 2D system is populated
(Solid arrows in figure \ref{zerofield}b).  Tunneling matrix element
effects reduce the tunnel current at high energies and densities,
causing the 2D band to appear as a peak rather than a
step\cite{Lebens87,Dial07}.  Because of the additional energy to
create the plasmon, creating a plasmaron requires more energy than
creating an ordinary hole.  More energetic holes occur at more
negative energies in our spectra.  Thus, the plasmaron appears as a second
edge below the 2D band edge in the spectrum (dashed arrows in
\ref{zerofield}b).  Both edges can be emphasized by differentiating
the data with respect to tunnel voltage to provide $d^2 I / dV^2$ and
smoothing it by convolving with a $\sigma=190\mathrm{\mu{}eV}$
Gaussian to remove resulting high frequency noise (figure
\ref{zerofield}d).

We identify the edges of the 2D and plasmaron band edges by the
location of the peak in $d^2I/dV^2$ (figure \ref{zerofield}c).  At
densities above $5\times10^{10}$, the 2D band edge lies at
$E_{2d}=(-0.403 \pm 0.003) meV \times N_{2d}/(10^{10}\mathrm{cm^{-2}})
+ (0.2 \pm 0.01) meV$, where $N_{2d}$ is the electron density of the
2DES.  The offset is due to our choice of the peak in the derivative
of the TDOS as our band edge.  For a non-interacting system, $m^*=\pi
\hbar^2/(dE_{2d}/dN_{2d})$, allowing us to extract $m^*=0.060\pm0.001
m_0$, where the error is dominated by systematic errors in the
calibration of the lever arm.  This is slightly smaller than the bulk
value of $m^*=0.063$ assumed in the energy calibration, implying a
somewhat larger energy bandwidth than expected.  Note this bandwidth
is the energy difference between suddenly creating a hole at the
bottom of the band and suddenly creating one at the Fermi energy, and
it is not in general equal to the chemical potential.  While the details
giving rise to this larger bandwidth are complex, we note that both
band-nonparabolicity\cite{Kane57} and interaction
effects\cite{Jalabert89a} are expected to increase this bandwidth
slightly.

We find a good empirical fit for the location of the plasmaron edge to
be $E_{pl} = E_{2d} - (1.53 \pm 0.02 \mathrm{meV}) \times
\sqrt{N_{2d}/(10^{10}\mathrm{cm^{-2}})} - 0.2 \pm 0.01 \mathrm{meV}$.
While this density dependence has not been predicted in the
literature, we note the overall $\sqrt{N_{2d}}$ is suggestive of the
density dependence of the plasmon component of the plasmaron, with
$\hbar \omega_p(k) = \sqrt{n e k/ ( 2 m \epsilon)}$.

\begin{figure}
\resizebox{5in}{!}{\includegraphics{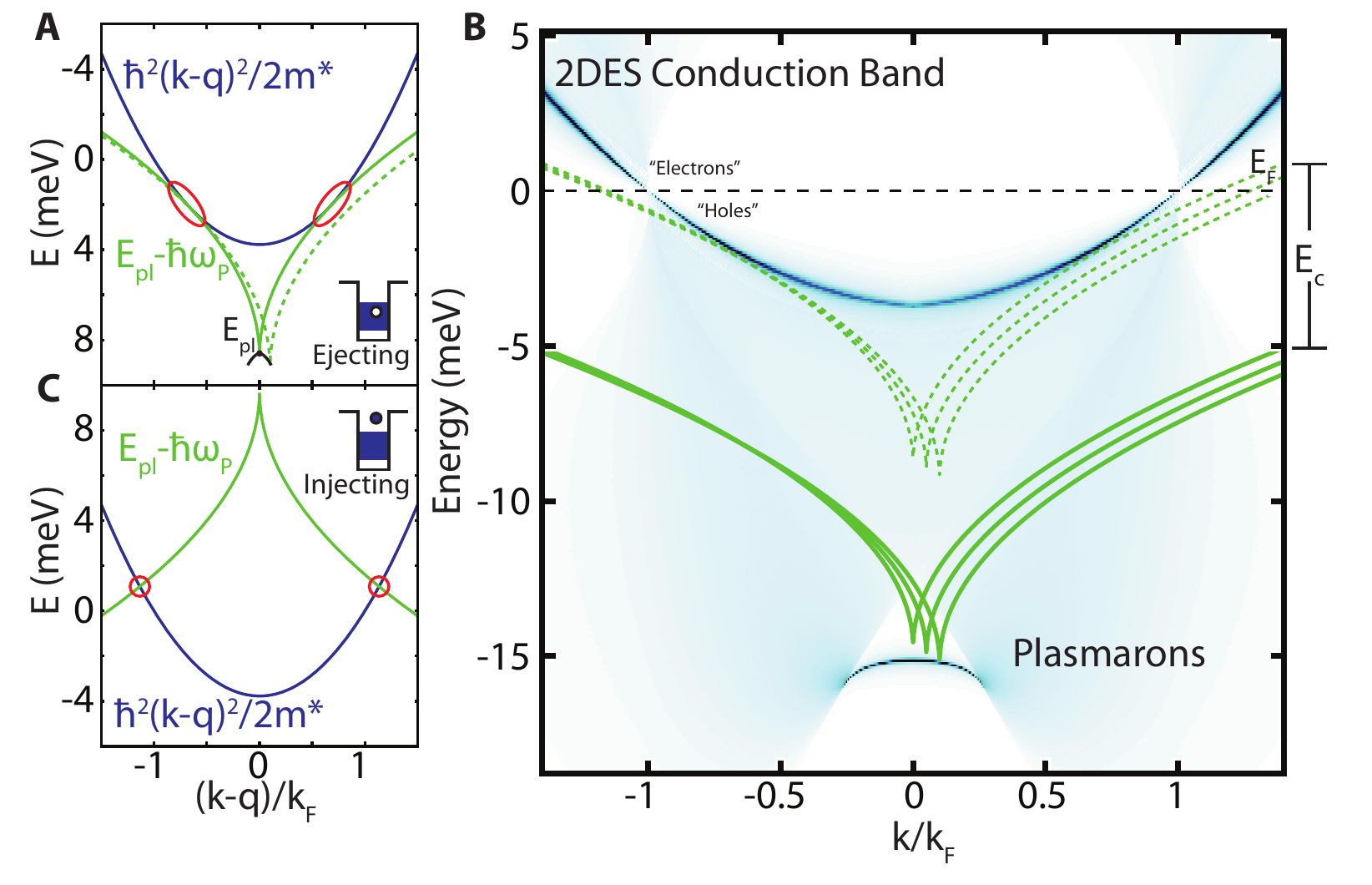}}
\caption{\label{plasmondisp}A simple cartoon model for the plasmaron
  can be developed by considering the resonance condition for the
  dispersion curves.  In A, the tangency between the hole (blue) and
  plasmon (green) dispersion results in a large phase-space
  enhancement for hole-plasmon coupling (red circles), resulting in
  the long-lived plasmaron (black line).  The $y$-axis has been drawn
  with more energetic holes downward to match the sign convention in
  our spectrum.  A more accurate RPA calculation of the spectral
  function $A(k,\omega)$, roughly the momentum resolved SPDOS, is
  shown as a colorscale plot in (B). Dark blue peaks
  corresponding to the normal electron/hole and the plasmaron are
  visible.  The general behaviour is consistent with the ``cartoon''
  model (solid blue, green lines) if a fixed coupling energy $E_C$ of
  about 5 meV is included (dashed green lines).  If this argument is
  repeated for electrons (C), the dispersions cross at a sharp angle
  rather than tangentially, and there is no phase-space enhancement.
  In C, more energetic electrons are plotted upwards to match the sign
  convention in our spectrum.}
\end{figure}

While detailed calculations of the plasmaron energy and lifetime exist
elsewhere
\cite{Hedin67,Lundqvist67,Bergersen73,Jalabert89a,Jalabert89b,vonAllmen92,Guillemot93},
a simple ``cartoon'' model aids in developing intuition and in
understanding how the heterostructure can modify the plasmaron
structure.  The plasmaron exists because, for a small range of
energies and wavevectors, it is possible to create a composite excitation
with wavevector $k$ consisting of resonantly bound holes of wavevector $k-q$ and plasmons
of wavevector $q$.  To create a plasmaron with energy $E_{pl}(k)$, this
resonance condition is given by
$$E_{pl}(k)=\frac{\hbar^2 \left[ k_f^2-(k-q)^2 \right]}{2m^*} + \hbar
\omega_p(q) + E_c(q,k)$$ Here $\hbar \omega_p(q)$ is the energy to
create a plasmon of wavevector $q$ (proportional to $\sqrt{N_{2d} q}$
at small $q$\cite{Stern67}) and $E_c$ is a coupling energy.  If we
momentarily neglect the coupling energy we can graphically solve for
resonance by rearranging this as $E_{pl}(k)-\hbar
\omega_p(q)=\frac{\hbar^2\left[k_f^2-(k-q)^2\right]}{2m^*}$ (Figure
\ref{plasmondisp}A; note the energy axis is inverted to match the sign
convention in our experimental data).  On doing so, it becomes clear
that for some values of $E_{pl}(k)$, a resonance occurs not for a
single value of $q$ but instead across a range of $q$ vectors where
the hole (blue) and plasmon (green) dispersion become tangent (red
circles); the hole-plasmon coupling is strongest when the hole and
plasmon group velocities match.  This results in a strong resonance
that allows the coherent screening of the injected quasiparticle by a
cloud of real plasmons, creating a long-lived excitation at $E_{pl}$.
As the momentum of the plasmaron $k$ is increased (black line),
$E_{pl}$ must increase to keep the plasmon and hole dispersion curves
tangent (dotted green line).  At the same time, the relevant plasmon
$q$ vector increases.  Because the coupling of the hole to the plasmon
is through the Coulomb interaction, it dies away as $1/q$.  This
weakening of the hole-plasmon coupling together with a decrease in the
resonant phase space destroys the plasmaron at large $k$.  For typical
densities in our device, this is predicted to occur at $k \sim 0.15
k_f$\cite{Jalabert89a}.  The plasmaron dispersion curve and its range
are pictured in black as the small upside-down u near the bottom of
Figure 2a.

Because the plasmon dispersion curve becomes steeper as
$\sqrt{N_{2d}}$, the plasmaron becomes more widely seperated from the
2D band as the density is increased.  At the same time, the relevant
$q$-vector increases, weakening and ultimately destroying the
plasmaron as the 2DES density is increased.

The plasmaron, then, is a long lived excitation that only exists at
small wave-vectors and small electron densities.  It is made up of
resonantly coupled holes and plasmons of equal but opposite
wavevectors centered about the $k$ vector of the plasmaron.  The
wavevector of the plasmon and hole involved is large: $\sim 0.66\ k_f$
for a GaAs 2DES at $N_{2d}=1\times10^{11}$.  A comparison of the
general form of the plasmaron dispersion from this argument to the
Random Phase Approximation (RPA) spectral function $A(k,\omega)$ (essentially
the momentum resolved SPDOS) calculated as per [\citen{Jalabert89a}] is included in Figure
\ref{plasmondisp}B.

In the case of injecting electrons into the quantum well, larger
momenta $k-q$ gives higher energy excitations rather than lower.  When
repeating the above discussion the parabolic electron dispersion is
flipped with respect to the plasmon dispersion.  The plasmon and
electron curve then cross at a sharp angle rather than touching
tangentially (Figure \ref{plasmondisp}C).  Accordingly, there is no
broad region of resonance, and no long lived plasmaron above the Fermi
energy.  This is in marked contrast to the case in graphene, where no electron-hole asymmetry exists\cite{Hwang08,Polini08}.

\begin{figure}
\resizebox{5in}{!}{\includegraphics{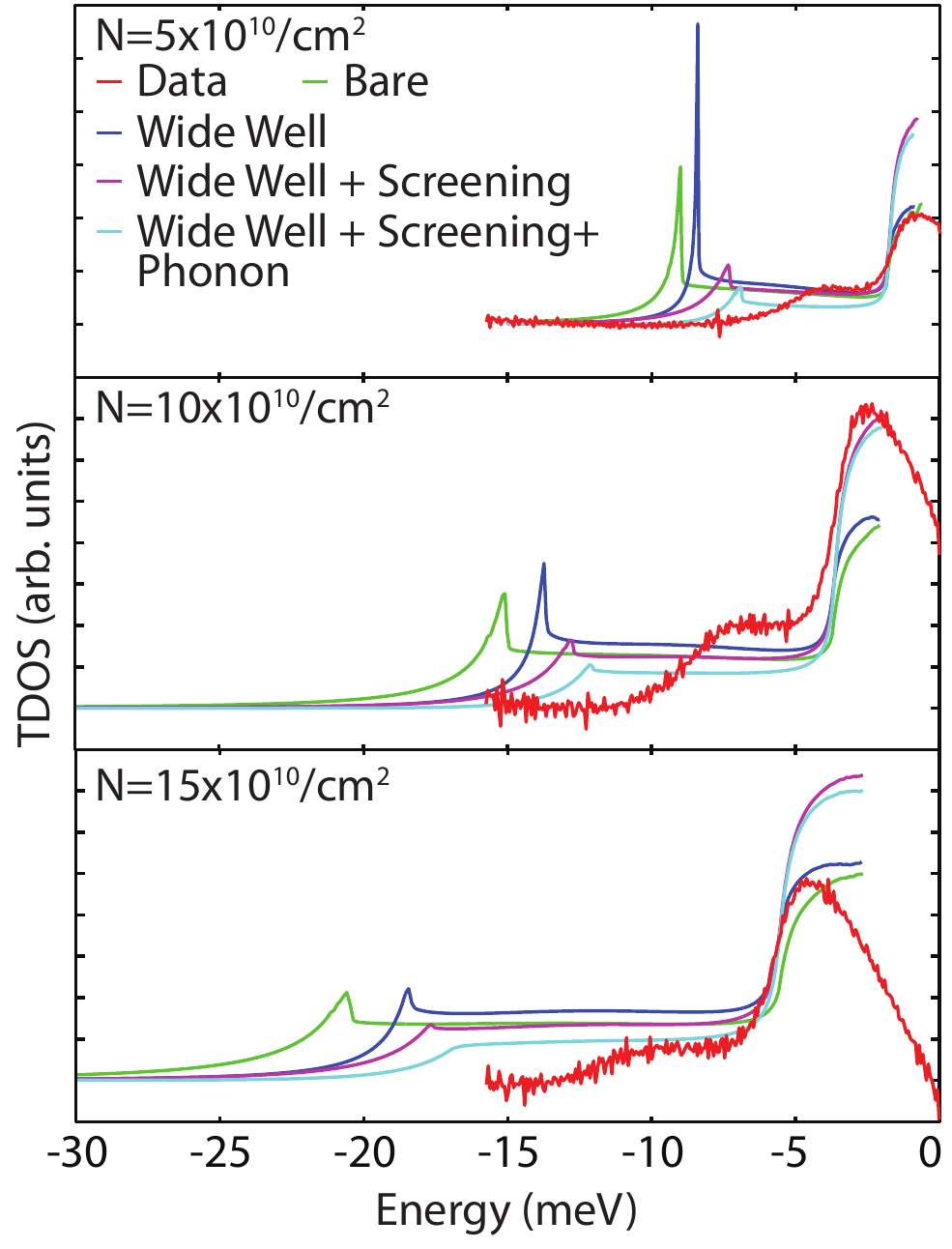}}
\caption{\label{RPA}Comparison of measured TDCS spectra to RPA
  calculations.  Up to an arbitrary scale factor in the TDOS axis and
  smooth distortions due to matrix element effects, the measured
  spectra can be quantitatively compared to theoretical calculations.
  Here, comparisons are made to a calculation with the bare Coloumb
  potential (Bare), a calculation in a 230\AA{} square quantum well
  (Wide Well), a calculation in a 230\AA{} well including screening
  from a nearby metallic electrode (Wide Well + Screening), and a
  calculation that also includes phonon coupling (Wide Well +
  Screening + Phonon).  All of these overestimate the coupling energy
  of the plasmaron and the size of the peak at the plasmaron edge, but
  as the Coulomb interaction is softened, the agreement between the
  calculation and the experiment grows better. }
\end{figure}

At any given wave-vector the plasmaron is sharply peaked at a
particular energy; however, when this spectrum is averaged across all
wave-vectors, the added structure due to the plasmaron is expected to
appear as a step, possibly with a peak at the low energy edge
depending on the lifetime of the plasmaron.  For an isolated quantum
well of infinitesimal thickness, this edge is predicted to lie from
two to six times the 2D Fermi energy below the 2D band edge at a
density of $1\times10^{11} cm^{-2}$, depending on what approximation
scheme is used in calculating the spectral function.  The plasmaron
step appears much closer to the Fermi energy in our data, separated
from the band edge by only roughly $1.3 \times E_F$ at $1\times10^{11}
cm^{-2}$.  The SPDOS calculated using RPA (as in Ref
[\citen{Jalabert89a}]) for an isolated infinitely thin well is
superimposed on measured spectra in figure \ref{RPA} at a variety of
densities, showing this discrepancy.  However, because the energy of
the plasmaron is extremely sensitive to both the plasmon dispersion
and the Coulomb interaction, a number of features in our structure not
present in the simplest calculations tend to considerably reduce the
distance between the 2D band edge and the plasmaron.  The 230\AA{}
wide square quantum well reduces the effective electron-electron
interaction at short distances, and can be accounted for by the
addition of a ``form factor'' to the Coulomb potential.  Doing so
moves the plasmaron edge somewhat closer to the band edge (figure 3).
In addition, the nearby metallic tunnel electrode screens the Coulomb
interaction at large distances and also tends to reduce the plasmon
energy at large wavevector\cite{Adolfo75,Hwang01}; this can also be
incorporated into the form factor.  This further reduces the
discrepancy as well as smoothing the plasmaron peak into more of a
step.  Finally, coupling to optical phonons modifies the dielectric
function somewhat at energies and wavevectors relevant to the
plasmaron, further reducing the plasmaron energy and completing the
transformation of the plasmaron contribution to the SPDOS to a step
rather than peak.

The calculated energy spectrum is still substantially
different from the measured spectrum.  RPA underestimates the
screening of the Coulomb interaction by the correlation hole around
injected quasiparticles; more accurate calculations would be expected to
further reduce the energy of the plasmaron.  The exquisite sensitivity
of the plasmaron feature to the electron-electron interactions make
it an excellent benchmark for testing approximate methods in
many-body theories.

\begin{figure}
\resizebox{6in}{!}{\includegraphics{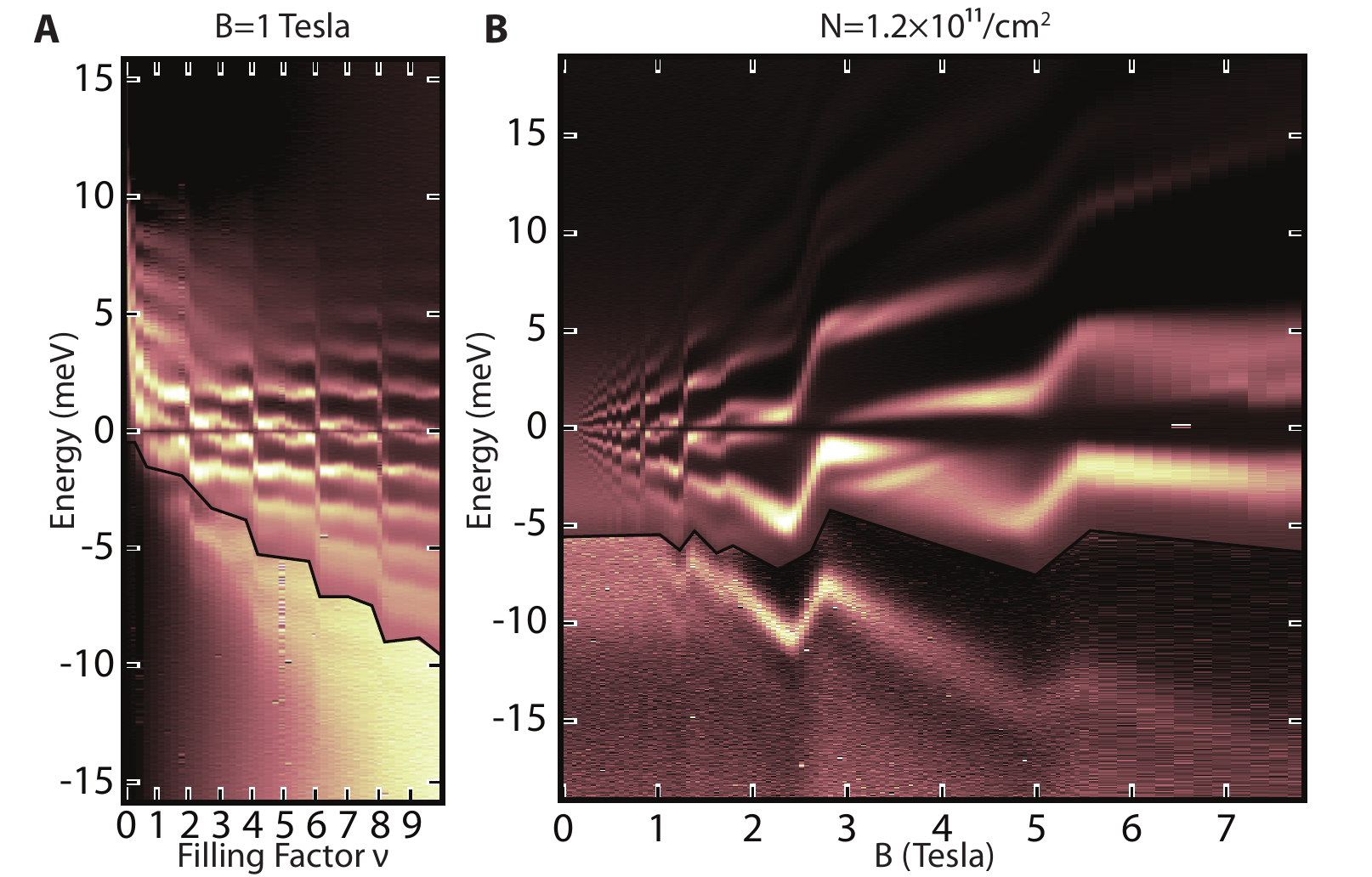}}
\caption{\label{fieldsweep} TDCS spectra showing the development of a
  magnetoplasmaron.  Sweeping density at a fixed field of 1 Tesla (A),
  the ``magnetoplasmaron'' appears as a series of ``ghost'' Landau
  levels $\hbar \omega_c$ below the $N=0$ Landau level.  On holding
  the density fixed and sweeping the magnetic field, the separation
  between the ``magnetoplasmaron'' and the $N=0$ Landau level is seen
  to vary linearly with B as $\hbar \omega_c$, and can be seen to
  merge into the $B=0$ plasmaron band.  The contrast has been enhanced
  below the black line just below the 2D band edge to show the
  plasmaron feature more clearly.  }

\end{figure}

On applying a quantizing magnetic field, we find the plasmaron
step breaks up into a series of faint ``ghost'' Landau levels below the 2D
band edge (Figure \ref{fieldsweep}A).  We note that the 2D density is
measured independently using magnetocapacitance in this measurement,
confirming that there is no offset on the density axis of this
spectrum and thus confirming our identification of the N=0 Landau level.
On applying a quantizing magnetic field, the energy of
creating a hole becomes non-dispersive, discretized by Landau
quantization into flat bands separated by $\hbar \omega_c$ .  At the
same time, the dispersion curve of magnetoplasmons is gapped by the
cyclotron energy, is rather flat, and has one or more magneto-roton
extrema at long wavelengths\cite{Kallin84,Kallin85}.  Thus, repeating
the ``cartoon'' arguments applied to the zero-field plasmon, it seems
reasonable for the band of plasmarons responsible for the plateau in
our data to sharpen into one or more ``ghost'' Landau levels lying
below $N=0$, which we label ``magnetoplasmarons'' by analogy to
magnetoplasmons.  This is indeed observed (figure \ref{fieldsweep}B);
  on varying the magnetic field while holding the bias voltage fixed
  (roughly fixing the density), a number of ``ghost'' Landau levels
  can be observed, with separations that scale with the cyclotron
  energy.  In particular, at $\nu=4$ at 1 Tesla, two ``magnetoplasmaron''
  peaks are visible below the $N=0$ Landau level with an inter-peak
  splitting of $1.9 \pm 0.1$ meV, similar to the cyclotron energy
  $\hbar \omega_c$ of $1.7$ meV at this field.  The splitting between
  the N=0 Landau level and the first plasmaron peak is complicated by
  the exchange splitting of the bottom Landau level; however, taking
  the energy of the $N=0$ level to be the mean energy of the spin up
  and spin down peaks, the measured splitting is $2.4 \pm 0.1$ meV,
  significantly larger than the cyclotron energy.

Once the magnetic field is applied, we chiefly observe magnetoplasmaron
features within the band of energies occupied by the plasmaron plateau
at zero field.  The magnetoplasmaron peaks sharpen and move away from the
$N=0$ Landau level as the cyclotron energy grows, but they largely
vanish as they fall below the energy of the plasmaron edge at zero
magnetic field.  The exact mechanism of this cutoff at high magnetic
fields is currently unknown.  However, the polarizability of the 2DES
at high magnetic field has a similar overall envelope to that at zero
magnetic field\cite{Roldan10}; this may be responsible for the similar
cutoff energies and densities.

Due to the accumulated expertise in fabricating nano-scale structures
in GaAs, the existence of plasmarons in this system may lead to new
applications and research lines.  Normally high energy quasiparticles
in Fermi liquids are short-lived due to decay by particle-hole
creation.  In contrast, plasmarons are predicted to be long-lived,
creating the possibility of accurate high energy spectroscopy and
barrier transmission measurements. Furthermore, the plasmaron carries
charge and can in principle easily be manipulated and focused in GaAs
using electrostatic gates\cite{Spector90,Sivan90}.  Finally, the
plasmon component of the plasmaron can strongly couple it to
terahertz light, creating the possibility of novel electro-optical
devices.

\bibliography{main}

\newcommand{\noopsort}[1]{} \newcommand{\printfirst}[2]{#1}
  \newcommand{\singleletter}[1]{#1} \newcommand{\switchargs}[2]{#2#1}
\begin{thebibliography}{10}

\bibitem{Hedin67}
L.~Hedin, B.~Lundqvist, S.~Lunqvist, {\it Solid State Communications\/} {\bf
  5}, 237 (1967).

\bibitem{Lundqvist67}
B.~I. Lundqvist, {\it Zeitschrift f\"ur Physik B Condensed Matter\/} {\bf 6},
  193 (1967).

\bibitem{Tediosi07}
R.~Tediosi, N.~P. Armitage, E.~Giannini, D.~van~der Marel, {\it Phys. Rev.
  Lett.\/} {\bf 99}, 016406 (2007).

\bibitem{Bergersen73}
B.~Bergersen, F.~Kus, C.~Bolmberg, {\it Can. J. Phys.\/} {\bf 51}, 102 (1973).

\bibitem{vonAllmen92}
P.~von Allmen, {\it Phys. Rev. B\/} {\bf 46}, 13345 (1992).

\bibitem{Jalabert89a}
R.~Jalabert, S.~Das~Sarma, {\it Phys. Rev. B\/} {\bf 39}, 5542 (1989).

\bibitem{Jalabert89b}
R.~Jalabert, S.~Das~Sarma, {\it Phys. Rev. B\/} {\bf 40}, 9723 (1989).

\bibitem{Hwang08}
E.~H. Hwang, S.~Das~Sarma, {\it Phys. Rev. B\/} {\bf 77}, 081412 (2008).

\bibitem{Polini08}
M.~Polini, {\it et~al.\/}, {\it Phys. Rev. B\/} {\bf 77}, 081411 (2008).

\bibitem{Bostwick10}
A.~Bostwick, {\it et~al.\/}, {\it Science\/} {\bf 328}, 999 (2010).

\bibitem{Chan97}
H.~B. Chan, P.~I. Glicofridis, R.~C. Ashoori, M.~R. Melloch, {\it Phys. Rev.
  Lett.\/} {\bf 79}, 2867 (1997).

\bibitem{Dial07}
O.~E. Dial, R.~C. Ashoori, L.~N. Pfeiffer, K.~W. West, {\it Nature\/} {\bf
  448}, 176 (2007).

\bibitem{Dial10}
O.~E. Dial, R.~C. Ashoori, L.~N. Pfeiffer, K.~W. West, {\it Nature\/} {\bf
  464}, 566 (2010).

\bibitem{Lebens87}
J.~A. Lebens, R.~H. Silsbee, S.~L. Wright, {\it Appl. Phys. Lett.\/} {\bf 51},
  840 (1987).

\bibitem{Kane57}
E.~O. Kane, {\it Journal of Physics and Chemistry of Solids\/} {\bf 1}, 249
  (1957).

\bibitem{Guillemot93}
C.~Guillemot, F.~Cl\'erot, {\it Phys. Rev. B\/} {\bf 47}, 7227 (1993).

\bibitem{Stern67}
F.~Stern, {\it Phys. Rev. Lett.\/} {\bf 18}, 546 (1967).

\bibitem{Adolfo75}
A.~Eguiluz, T.~K. Lee, J.~J. Quinn, K.~W. Chiu, {\it Phys. Rev. B\/} {\bf 11},
  4989 (1975).

\bibitem{Hwang01}
E.~H. Hwang, S.~Das~Sarma, {\it Phys. Rev. B\/} {\bf 64}, 165409 (2001).

\bibitem{Kallin84}
C.~Kallin, B.~I. Halperin, {\it Phys. Rev. B\/} {\bf 30}, 5655 (1984).

\bibitem{Kallin85}
C.~Kallin, B.~I. Halperin, {\it Phys. Rev. B\/} {\bf 31}, 3635 (1985).

\bibitem{Roldan10}
R.~Rold\'an, M.~O. Goerbig, J.-N. Fuchs, {\it Semiconductor Science and
  Technology\/} {\bf 25}, 034005 (2010).

\bibitem{Spector90}
J.~Spector, H.~Stormer, K.~Baldwin, L.~Pfeiffer, K.~West, {\it Appl. Phys.
  Lett.\/} {\bf 56}, 1290 (1990).

\bibitem{Sivan90}
U.~Sivan, M.~Heiblum, C.~P. Umbach, H.~Shtrikman, {\it Phys. Rev. B\/} {\bf
  41}, 7937 (1990).

\end{thebibliography}

\bibliographystyle{Science}

\begin{scilastnote}
\item O.E.D. built the apparatus and performed measurement and analysis.  R.C.A. supervised the work and performed analysis.  O.E.D. and R.C.A. prepared the manuscript.  L.N.P. and K.W.W. performed the crystal growth.  This work was sponsored by the Basic Energy Sciences Program of the Office of Science of the US Department of Energy, contract number FG02-08ER46514.
\end{scilastnote}


\clearpage

\end{document}